\documentclass[runningheads]{llncs}
\usepackage[T1]{fontenc}
\usepackage{graphicx}
\usepackage{hyperref} %
\usepackage{multirow} %
\begin{document}
\title{Transversal PACS Browser API: Addressing Interoperability Challenges in Medical Imaging Systems}
\titlerunning{Transversal PACS Browser API}

\author{Diogo Lameira \inst{1} \and
Filipa Ferraz \inst{1}\orcidID{0000-0002-6912-2853}}
\authorrunning{D. Lameira et al.}

\institute{ALGORITMI Research Centre/LASI, University of Minho, Braga, 4710-057, Portugal\\
\email{pg50332@alunos.uminho.pt}\\
\email{filipa.ferraz@di.uminho.pt}}

\maketitle

\begin{abstract}
Advances in imaging technologies have revolutionised the medical imaging and healthcare sectors, leading to the widespread adoption of PACS for the storage, retrieval, and communication of medical images. Although these systems have improved operational efficiency, significant challenges remain in effectively retrieving DICOM images, which are essential for diagnosis and overall patient care. Moreover, issues such as fragmented systems, interoperability barriers, and complex user interfaces can often prevent healthcare professionals from efficiently accessing medical images. Addressing these challenges, the \textit{Transversal PACS Browser API} is a robust and user-friendly solution designed to enhance the process of querying and retrieving DICOM images. It offers advanced filtering capabilities through a variety of filter options as well as a custom field search, that allows users to easily navigate through large medical image collections with ease. Additionally, the application provides a unified interface for querying and retrieving from multiple PACS stations, addressing the challenges of fragmentation and complexity associated with accessing medical images. Other key features include the ability to pre-view images directly within the application. All of this contributes to the transversal nature of the API, serving not only healthcare providers, but anyone who relies on efficient access to these resources. To validate the performance and usability of the application, comprehensive testing was carried out with stakeholders of the field, the results of which showed general satisfaction, highlighting the API's clean design, ease of use, and effective search capabilities of the API, as well as the usefulness of previewing images within the application.

\end{abstract}

\section{Introduction}
In recent years, the medical imaging and healthcare fields have undergone substantial changes with the transition from analogue to digital imaging technologies. The digitalisation process in these sectors has resulted in the pervasive adoption of Picture Archiving and Communication Systems (PACS), which are indispensable for the storage, retrieval, and communication of medical images in digital format. Such systems contribute not only to more efficient operations for medical staff, but also facilitate the exchange of patient information across different institutions. It is, therefore, essential that images are managed efficiently and that they are accessible in order to facilitate diagnosis, treatment and overall patient care \cite{bib1}.

Despite the considerable advancements witnessed in the domain of PACS in recent times, there persists a significant challenge in ensuring the efficient retrieval of DICOM (Digital Imaging and Communications in Medicine) images from a multitude of medical imaging modalities and healthcare institutions. The healthcare sector is characterised by fragmented systems, interoperability issues and complex interfaces that impede the ability of healthcare providers, radiologists and other medical professionals to rapidly access the necessary medical images for patient care. Furthermore, the ever-increasing volume of medical images generated daily underscores the growing need for a sophisticated and user-friendly PACS browser application.

Taking this into consideration, the main motivation of this project is to address these challenges and contribute to the improvement of PACS systems by developing a robust and user-friendly application that serves as a PACS browser that will allow the query and retrieval of DICOM images from multiple stations, with several pre-defined and open fields filters while prioritizing specific features, like usability, efficiency, and adaptability to meet the evolving needs of healthcare providers.

The application aims to provide an efficient platform for managing and accessing medical images in a healthcare facility, for example, simplifying the often-complex task of navigating extensive collections of medical images. Users should be able to easily locate specific images, thus saving valuable time and improving work-flow efficiency. Additionally, one of the main ambitions of this application is its robust DICOM-compliant query and retrieval capability, which ensures seamless integration with various medical imaging systems.  Moreover, the PACS browser application prioritizes data security and privacy, which is a major concern in the healthcare sector, through a secure environment for the retrieval and visualization of sensitive data, ensuring the protection of patient information.

Furthermore, this application also aims to cater to the need for searching and managing medical images beyond healthcare facilities, serving as a tool for Information Technology (IT) professionals or researchers who rely on efficient access to these resources. 

\subsection{Investigation Methodology}
The literature review was conducted following the PRISMA (Preferred Reporting Items for Systematic reviews and Meta-Analyses) methodology to provide a systematic and organized approach to analysing existing research \cite{bib2}. This process aimed to investigate the current state of research related to PACS and their limitations, the intricacies of DICOM handling, and the methodologies employed in query and retrieve operations. Additionally, the review required to explore the integration of DICOM libraries and medical image retrieval systems to identify gaps and opportunities for enhancing existing processes. To achieve this, multiple academic databases were utilized, including PubMed, Scopus, and IEEE Xplore, employing various keyword combinations such as \textit{"PACS"}, \textit{"DICOM"}, \textit{"Medical Image Retrieval"}, \textit{"Query and Retrieve"}, \textit{"DICOM Integration Libraries"}, and \textit{"Limitations of PACS"}. This targeted search strategy ensured a thorough examination of the relevant literature, enabling the identification of key challenges and advancements in the field.

\subsection{PACS and Current Limitations}
PACS is a crucial component of modern medical imaging, enabling the acquisition, storage, communication, display, and management of diagnostic imaging studies. These systems have significantly improved healthcare efficiency, with studies indicating earlier patient discharges and reduced hospital stays in institutions using PACS \cite{bib3}. However, there is a notable lack of recent literature on PACS, underscoring the need for updated information on advancements and best practices.

Historically, conventional radiography relied on physical film processing, which was time-consuming, prone to data loss, and delayed clinical decision-making \cite{bib4}. In contrast, PACS streamlines the process by digitizing images, making them immediately accessible across networks. This transformation has enhanced the speed and convenience of image retrieval, interpretation, and sharing, offering features like multi-user access, on-demand availability, and advanced image processing, all of which contribute to improved patient care and operational efficiency.

PACS typically relies on DICOM standards and consists of four main components: devices to acquire images, communication networks, integrated visualization workstations, and PACS archive and server. Image acquisition devices include imaging modality devices and gateway computers, which convert proprietary formats into DICOM and may perform basic image processing. Images are obtained either by digitizing films or through direct digital acquisition with modern devices \cite{bib4,bib5}.

The communication network connects PACS components, including imaging devices, gateway computers, servers, and workstations, using networks like LANs or WANs to ensure seamless data transfer. The PACS archive server acts as the system's core, storing images and patient data received from acquisition gateways and hospital and radiology information systems. It includes short-term and long-term storage and performs functions like image routing, updating databases, and retrieving relevant studies. Workstations, or diagnostic workstations, are integral for primary diagnosis and replace traditional film review tools. Equipped with DICOM viewers, they enable advanced image analysis, manipulation, and annotation. The interaction between archives and workstations is typically managed through the query \& retrieve mechanism, ensuring efficient image access and evaluation \cite{bib6}.

Key questions regarding PACS performance include addressing its limitations and identifying strategies to enhance functionality. Major challenges include processing large amounts of medical image data, low bandwidth and slow communication networks, and inadequate diagnostic monitors. Studies have also highlighted issues like insufficient user training, difficulty finding images due to duplicate IDs and incomplete data, and the lack of desired features like remote access and mobile functionality. Additionally, concerns about PACS downtime and system reliability persist \cite{bib7,bib8,bib9}.

To overcome these challenges, strategies include improving technical infrastructure, such as allocating higher bandwidth and ensuring faster networks. Training programs for users can address knowledge gaps, while implementing a unique patient registry can resolve issues with data redundancy and image loss. Investigating and addressing the causes of PACS downtime can further enhance system reliability and overall performance.

\subsection{Existent Technologies}
Recent technologies have emerged to support healthcare providers in navigating repositories of patient data, addressing the need for efficient medical imaging workflows. These tools often include built-in query and retrieval functionalities in medical image viewers like RadiAnt, MicroDicom, and Pacsbin, which integrate directly with PACS to streamline access to medical images. This built-in capability is crucial, enabling healthcare professionals to locate, retrieve, and manage images directly from PACS systems.

One notable example is Semantic DICOM (SeDI), which incorporates semantic search capabilities into PACS, enabling users to query images based on specific attributes, such as tumour size or other diagnostic criteria. This advanced functionality supports clinicians by integrating analytic results with patient data, optimizing diagnosis and treatment processes, and providing access to reference cases. Its ability to handle complex, attribute-based searches makes it a powerful tool for enhancing decision-making \cite{bib10}. Another example is XNAT, an imaging informatics platform with a DICOM \textit{Query-Retrieve} plugin. This plugin allows targeted queries, batch processing, and bidirectional data transfer between PACS systems and XNAT\footnote{XNAT is an open-source, extensible imaging computer software platform for image-based research.}. Its tailored data organization features include relabelling and structured session management. However, the plugin has limitations regarding data security, as sensitive protected health information is sometimes included in XML documents during the retrieval process, presenting potential privacy concerns \cite{bib11}.

\section{Methods and Materials}
The application development was carried through the Qt Creator integrated development environment based on Qt 6.6.0 and the Clang 13.0 compiler. Together, they enable multi-platform app development in C++, while also leveraging Qt’s libraries for UI design, networking, and data management. It aims to offer a straightforward interface and robust backend, providing users with the ability to connect to PACS servers, execute searches based on various criteria, retrieve and store specific imaging studies or series locally, as well previewing them. Therefore, the application is structured around these functionalities and fundamental DICOM services, which ensure compliance with medical imaging standards, offering essential functions such as query, retrieve, store, and server connection verification.

In terms of implementation, Qt's GUI toolkit facilitated the creation of a responsive and intuitive interface, through a signal-slot mechanism that enabled seamless interaction between user inputs and backend operations. This is a fundamental feature that facilitates seamless communication between different components within a Qt application. At its core, this mechanism enables objects to emit signals in response to certain events or changes in their state, and other objects can listen for these signals and react accordingly by executing designated slots. One of the key advantages of Qt's signal-slot mechanism is its flexibility and simplicity, allowing for loosely coupled communication between different parts of the application, and, at the same time, promoting modular and maintainable code. Objects do not need to have direct knowledge of each other to communicate; instead, they can interact through signals and slots, making it easier to manage dependencies and modify the application's behaviour without impacting unrelated components. Therefore, when a signal is emitted, Qt's event loop mechanism dispatches the signal to all connected slots, invoking their corresponding functions, which is particularly useful in GUI programming, where user interactions trigger events that need to be handled by different parts of the application. For example, a button click event can be connected to a slot that performs a specific action, such as updating the UI or processing data.

The application follows a well-structured architecture that comprehends multiple essential components, designed to facilitate the querying, retrieval, and management of medical images from PACS servers. The programming language C++ forms the backbone of the application, handling the core logic, while the Qt framework provides a robust platform for the entire development process, allowing for seamless interaction across various operating systems, streamlining platform-specific code and ensuring consistent user experience.

\begin{figure}[h!]
\centering
\includegraphics[width=0.9\textwidth]{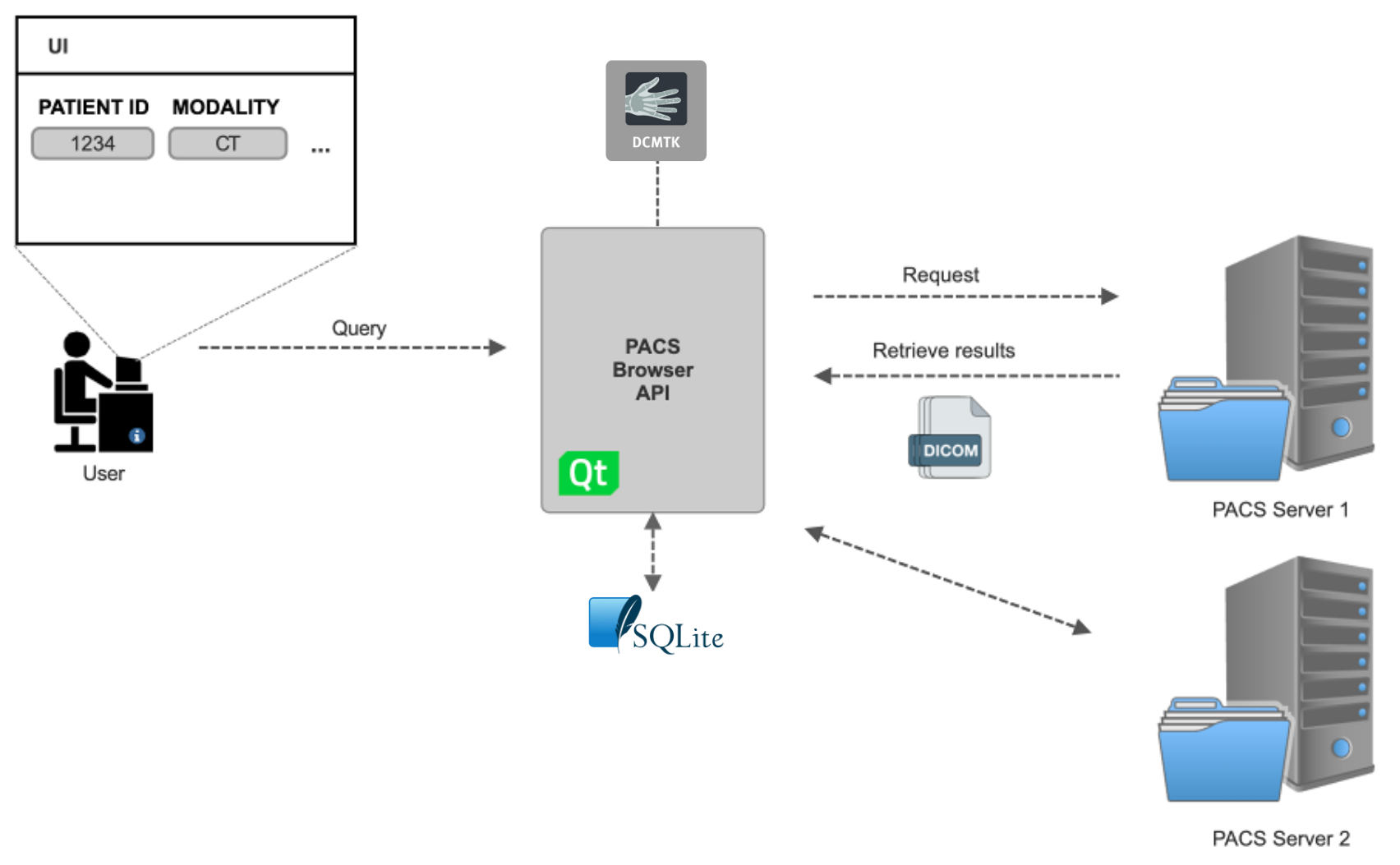}
\caption{Proposed application's architecture [by author].}\label{fig1}
\end{figure}

Regarding data management and medical image transmission, the application relies on the DICOM protocol, which is essential for communicating with PACS servers in a secure, standardized way. Therefore, the architecture of the application is centred around enabling efficient retrieval of medical images, following a client-server model, represented by Service Class User (SCU) and Service Class Provider (SCP) in this case, where the frontend was built using Qt’s GUI elements and the backend handles communication with multiple PACS servers, leveraging Qt's networking capabilities to manage communications and coordinate the retrieval of query results. The backend can efficiently coordinate requests to the PACS servers, retrieving and aggregating the results. Additionally, it is essential to incorporate an authentication mechanism to secure user access and maintain data integrity. When it comes to storing important data such as user credentials, that is accomplished with the support of a local SQLite database, which provides lightweight and reliable data management. To maintain user preferences and configuration settings between sessions, the application employs Qt's \textit{QSettings} mechanism, enabling a consistent experience by storing user preferences locally in a straightforward, platform-independent way. Furthermore, DICOM ToolKit (DCMTK) serves an important role in the application, providing tools and libraries to manage the complexities of the DICOM protocol and facilitating smooth integration with PACS servers \cite{bib12}. Figure~\ref{fig1} visually illustrates the architecture of the application, providing an overview of its structural components and their interconnections.

\section{Transversal PACS Browser API}
The proposed solution, the \textit{Transversal PACS Browser API}, aims to provide an integrated, efficient, and user-friendly tool for querying and retrieving DICOM images from multiple PACS stations. Hence, the application is grounded in the principles of interoperability, usability, and efficiency, aiming to align with the current needs of healthcare providers to improve image retrieval processes and overall workflow efficiency in radiology and other medical imaging-dependent specialities. Therefore, the key innovation of this application lies in its transversal nature. The application was designed to overcome the limitations of current fragmented systems by allowing seamless access to medical images across different PACS stations. Through the implementation of a unified interface, the application abstracts the complexities of interacting with various systems, ensuring that users can query and retrieve images without the need to navigate different systems manually.

\subsection{Backend and User Interface}
The development of the application involved two fundamental aspects: the backend connections and the User Interface (UI) design development. The backend connections are the backbone of the application, enabling seamless communication between the user’s system and the PACS servers, by implementing the DICOM protocol for query and retrieval of medical images, ensuring compliance with industry standards. By establishing robust communication networks, the backend can handle high volumes of data while maintaining reliable connections to PACS servers. As for the UI design, it focuses on creating an intuitive and responsive experience for the application users. The complexity of managing large datasets, including performing search queries and retrieving medical images, must be presented in a way that minimizes cognitive burden and maximizes efficiency. This means designing an interface that allows for quick navigation and easy access to core functionalities to cater to various user preferences and workflows. Combining both sides of the API, the backend ensures that the application is capable of handling the technical complexities of connecting to diverse PACS systems, while the UI ensures that this complexity remains hidden from the user.

\subsection{User Flow and Application Features}
Starting by the main \textit{View/Controller}, it is a central component of the PACS browser API, responsible for managing the initial interface of the application. When users launch the application, they are greeted with an intuitive main window that provides them with a central button for logging in. 

\subsubsection{Authentication}
When a user clicks the \textit{Login} button, the UI transitions to the login page, where they can enter their credentials. This page includes input fields for the username and password, accompanied by a submission button. Moreover, a registration page was implemented within the application but remains hidden from the main view, as user account creation should be restricted to only authorized administrators, ensuring that standard users cannot create additional accounts independently. The login and registration processes in the application are important for safeguarding sensitive information. By requiring users to create an account and authenticate their identity, the application ensures that only authorized individuals can access functionalities that could expose sensitive data, which is crucial in fields like healthcare where the privacy regulations are rigorous. Additionally, this feature is important since users can save specific settings in between sessions, including stations’ information and other preferences that should remain private.

When a new account is being created, the process is only allowed for the \textit{admin} user, the users are guided through a straightforward interface where they are required to enter a username and password. If the username is available, the application securely hashes the password provided by the user, using SHA256 hash functions, and, therefore, enhancing the security of user credentials. With the hashed password ready, the application prepares to store the new user's information in an SQLite database, a relational database management system that operates directly on disk files, eliminating the need for a separate server.

While the password storage system in the application uses hashing to improve security, it has several limitations, especially when it comes to ensuring full user privacy. Hashing transforms a password into a unique, fixed-length string that represents the original input, so instead of storing plain-text, the application saves only hashed versions, which significantly reduces the risk of exposing passwords. However, among some others, one key limitation is that hash functions, while irreversible in theory, are vulnerable to brute-force and dictionary attacks. In these attacks, an attacker systematically hashes large numbers of potential passwords until they find one that matches a stored hash, which is particularly effective against weak or common passwords. Therefore, this algorithm was adopted to simulate a real authentication system, but it lacks the robustness that would be required in a ready-to-use application.

After getting through the authentication system, the users are presented with the landing page on the main view. It works as a menu for the application, allowing navigation between key functionalities like querying PACS servers and accessing the settings, coordinating the flow of the user’s interactions, and ensuring that all subsequent processes are correctly initiated and handled. When the user first clicks the \textit{Query \& Retrieve} button on the application's main window, the PACS \textit{Query} view is launched, initiating a connection verification process (\textit{C-ECHO}) to all PACS stations that are saved locally in the application's settings. This \textit{C-ECHO} operation, part of the DICOM protocol, is essential for confirming the availability and status of each PACS server, allowing the verification of which stations are currently accessible, and ensuring that users can direct their queries to active servers only. By performing a connection verification to all known stations, the application offers users the flexibility to either query a specific station or query all available stations simultaneously if multiple servers are accessible at the time.

\subsubsection{Query}
After successfully opening the \textit{Query} view, the user can use the presented fields in the UI to search for specific \textit{Study}, \textit{Patient}, and \textit{Series} attributes, essential for effective image retrieval. Key study parameters include \textit{Study Date}, \textit{Study Time}, \textit{Study ID}, \textit{Referring Physician's Name}, \textit{Accession Number}, and \textit{Study Instance UID}. Furthermore, patient-related fields comprehend \textit{Patient ID}, \textit{Patient Name}, \textit{Sex}, and \textit{Birth Date}, while for series information users can specify a \textit{Modality}, \textit{Series Instance UID}, and \textit{Series Number} to perform their search. These filters align very closely with DICOM’s required keys for query on each of these levels, which represent the minimal set of attributes that must be supported \cite{bib13}.

Additionally, to provide more flexibility in querying, the application includes a custom field feature that allows users to select from an extensive list of DICOM tags beyond the standard set provided and, consequently, allows them to direct their queries to retrieve more specific information that suits their specific needs. This customization enhances the search capabilities and allows users to interact with the PACS server with greater precision and control over the data they access. After the user inputs their desired search criteria and clicks the \textit{Search} button, the application triggers a \textit{C-FIND} operation. The query sent consists of an instance from \textit{DcmDataset}, which is a class in the DCMTK library, designed to handle DICOM data elements. It serves as a container for DICOM attributes, allowing them to manipulate these elements effectively. Therefore, each dataset becomes a container for specifying tags and values that describe the requested information, consisting of pairs of unique tags and their corresponding values.

Since DICOM supports hierarchical querying (e.g., studies, series, images), the dataset can be adjusted dynamically to accommodate these varying query scopes. Queries in a DICOM system are directed to each level because each level serves a unique purpose in the diagnostic and organizational workflow of medical imaging and each level contains different attributes that might be searched for. However, in the context of this application, performing queries at the three hierarchical levels is essential for accommodating the fields that users may want to search for, since they can belong to any of these layers.

In terms of UI design, to present the query results to the user, a \textit{Tree Widget} is specifically designed to present data in a tree-like structure and aligning with the nature of complex datasets such as DICOM images. Each study is represented as a parent node in the tree, with the study's unique identifier (\textit{Study Instance UID}) as the primary identifier. Under each study node, the associated series are added as child nodes, with each series containing detailed information like \textit{Series Instance UID} and other attributes, such as modality and description. Additionally, the application also creates an \textit{Actions} column, which is populated with three buttons that allow users to interact with the study or series by performing certain actions, such as retrieving, previewing the images, or opening the destination folder after retrieval. The \textit{Preview} and \textit{Open} buttons are disabled by default until the user retrieves anything. After querying, the user can choose to either retrieve a whole study or specific series.

\subsubsection{Retrieval}

Starting by the case where the user chooses to retrieve a whole study, the retrieval process begins with a user-driven action in the UI, where the user clicks the \textit{Retrieve} button associated with a study item in the results tree widget, which triggers the application to initiate a sequence of backend operations.

The implementation of DICOM services — specifically \textit{Echo}, \textit{Find}, \textit{Move}, and \textit{Store} — provides the core functionality for communication between medical imaging systems within the application, enabling interaction with the PACS servers.

In the implementation of these services, the DCMTK library was used, providing a comprehensive framework for implementing DICOM communication. Therefore, this setup ensures that the application can seamlessly manage DICOM objects across different networks. Also, the use of DCMTK ensures that these services are implemented with high fidelity to DICOM protocols, making the application compatible with a wide range of PACS systems in clinical environments.

Firstly, the application attempts to locate the selected study in the existing study responses, which are a list of results previously retrieved from the PACS during the query phase and, once the study is found, the system proceeds to request the retrieval of the series information associated with that study. To do so, it is used the \textit{Find SCU} service to query the PACS for all series that belong to the identified study. If the query is successful, the series responses are stored, and the system begins to process each series. For each series, the system retrieves the \textit{Series Instance UID}, a unique identifier for that specific series. It then moves to the next phase: retrieving the individual images that belong to this series. This is done by sending a request to the PACS for all images under the specified \textit{Series Instance UID} using the Find SCU. If the PACS responds with the requested images, the system iterates through each image, extracting its \textit{SOP Instance UID}, which is the unique identifier for each image in the DICOM standard. Each image's details are then prepared for retrieval. At this point, the system constructs the necessary dataset containing all the relevant identifiers (\textit{Study Instance UID}, \textit{Series Instance UID}, \textit{SOP Instance UID}) required to precisely request each image. Once the system has identified the study, series, and individual images it needs to retrieve, the \textit{C-MOVE} operation is initiated and the PACS begins transmitting the images. This operation instructs the PACS server to transfer the requested images to a specified destination, in this case a local storage server, and the system keeps track of how many images are expected, how many are successfully retrieved, and whether any errors occur during the transfer process. This operation is repeated for each image in the series, and for each series within the study. Moreover, the \textit{Store SCP} plays a critical role in receiving and storing the images on the local system. It acts as a server on the local machine, capable of accepting incoming images from the PACS during a \textit{C-MOVE} operation.

\begin{figure}[h!]
\centering
\includegraphics[width=0.9\textwidth]{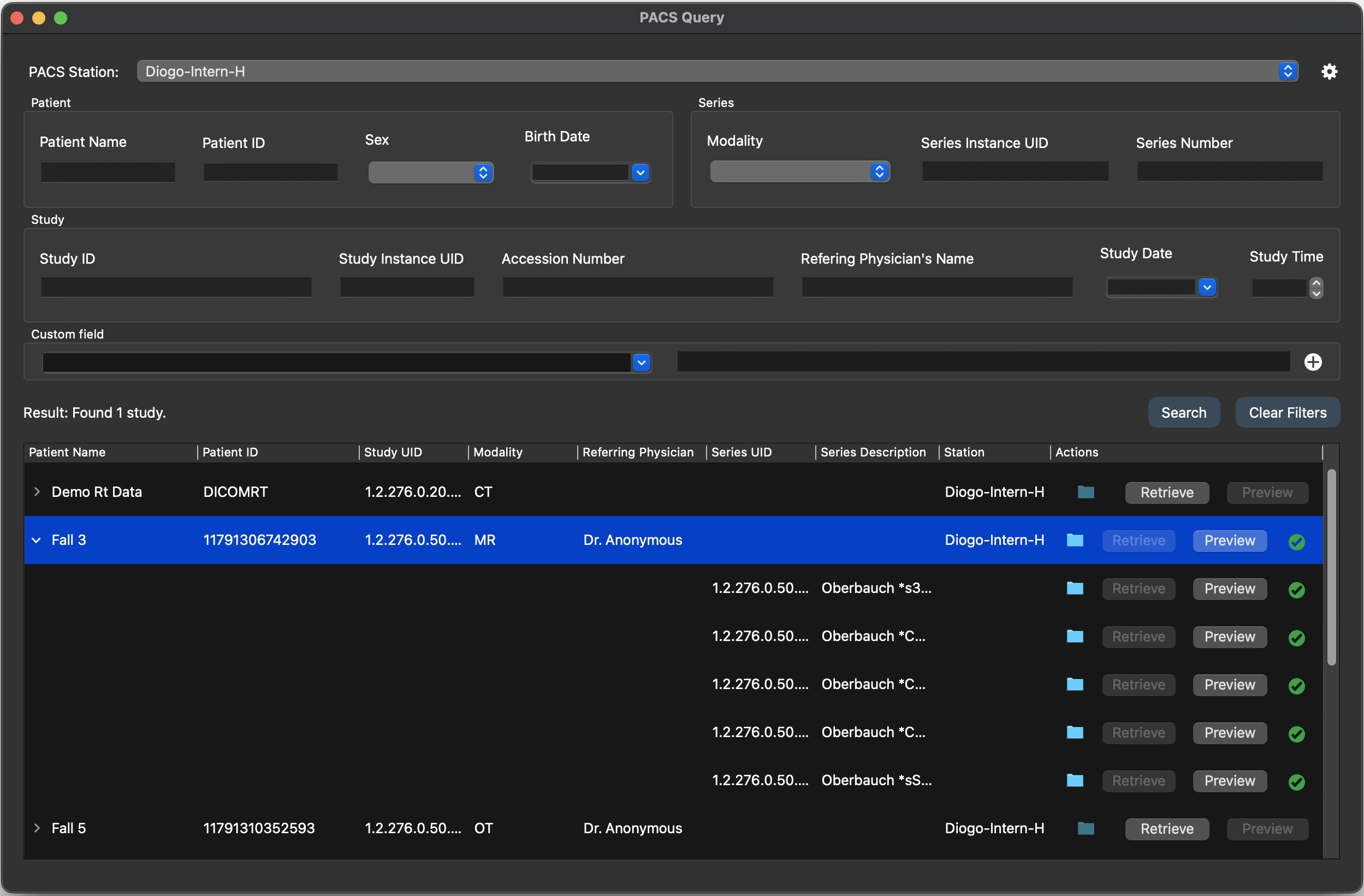}
\caption{Screenshot of the \textit{Query} view presenting an example of a successful study retrieval.}\label{fig2}
\end{figure}

In the event of a successful study retrieval, the system ensures that the user is informed that the retrieval has been completed, and further attempts to retrieve the same study are prevented by disabling the \textit{Retrieve} button associated with the study in the tree view. However, if the study retrieval fails, the primary objective is to keep the user informed about the failure while still allowing the retrieval to be retried by keeping the \textit{Retrieve} button enabled for the corresponding study. Additionally, to mark the failure visually, a cross-mark icon is displayed next to the study item. Figure~\ref{fig2} shows the \textit{Query} view presented to users upon a specific query and a successful study retrieval.

As for the case where the user chooses to retrieve a specific series, it works similarly to the study retrieval but focused on a specific series within a study. The system first gathers the selected series information from the UI – the \textit{Study} and \textit{Series Instance UIDs} associated with the chosen series. Additionally, the PACS station information is also retrieved from the results tree and complemented with the information stored in the saved settings. After that, the system sends a \textit{C-FIND} request at the series level to the station to which the selected series belong and consequently, a \textit{C-FIND} request at Image level for the images contained in the series. This process mirrors the study retrieval but is restricted to a specific series instead of the entire study. Once the images within the series are identified, the system prepares a \textit{C-MOVE} request for each image and uses the \textit{Store SCP} to store them.

\subsubsection{Image Preview}
After successfully retrieving a study or series, the user is presented with the option to preview the images. This preview process involves converting the images from the DICOM format to a standard, universally recognized image format, the JPEG format. Once converted, a dialogue window loads with the JPEG images, enabling users to quickly browse through them. This feature is designed primarily for rapid image inspection, allowing users to get a general sense of the study or series contents, without substituting the more advanced functionalities offered by traditional medical image viewers. While convenient, this preview lacks complex viewing tools typically needed for diagnostic tasks, such as adjusting image contrast and brightness, or applying specific windowing settings that help highlight subtle tissue or anatomical differences. Therefore, it serves as a practical tool for quick reference and validation of the retrieved images’ content and quality but is not intended to replace dedicated imaging software that clinicians rely on for detailed analysis and diagnostic accuracy.

The conversion process begins by sorting the DICOM images based on the \textit{Instance Number} tag to ensure they appear in the correct sequence. Each DICOM file is then processed individually, with its dataset loaded using the \textit{loadFile} method from the \textit{DcmFileFormat} class. This method extracts essential metadata, including the \textit{Transfer Syntax UID}, which specifies the compression format used. The appropriate codec is selected to decompress the image data, ensuring accurate interpretation of the pixel data.  Once decompressed, the pixel data is extracted and processed according to its bit depth (e.g., 8-bit or 16-bit) and type (monochrome or colour). Monochrome images are mapped to grayscale, while colour images use RGB mapping. Windowing techniques are applied to map the higher-bit-depth DICOM data (usually 12 or 16 bits) to 8-bit grayscale for display \cite{bib14}. If the DICOM file includes \textit{window width} and \textit{centre} parameters, these are used to optimize the image display; otherwise, default settings are applied. After processing, the pixel data is converted to a \textit{QImage} object, which supports JPEG output. The image is then saved in sequential filenames to maintain order and resources used for codec registration are released after processing each file.

The \textit{Preview} window allows users to navigate through images in a series using a slider below the display area. For studies with multiple series, additional navigation buttons let users switch between series within the same interface. Additionally, the window adjusts dynamically to the image dimensions, ensuring they fit the screen while preserving the original aspect ratio. An example of a specific study preview is presented below in Figure~\ref{fig3}.

\begin{figure}[h]
\centering
\includegraphics[width=0.5\textwidth]{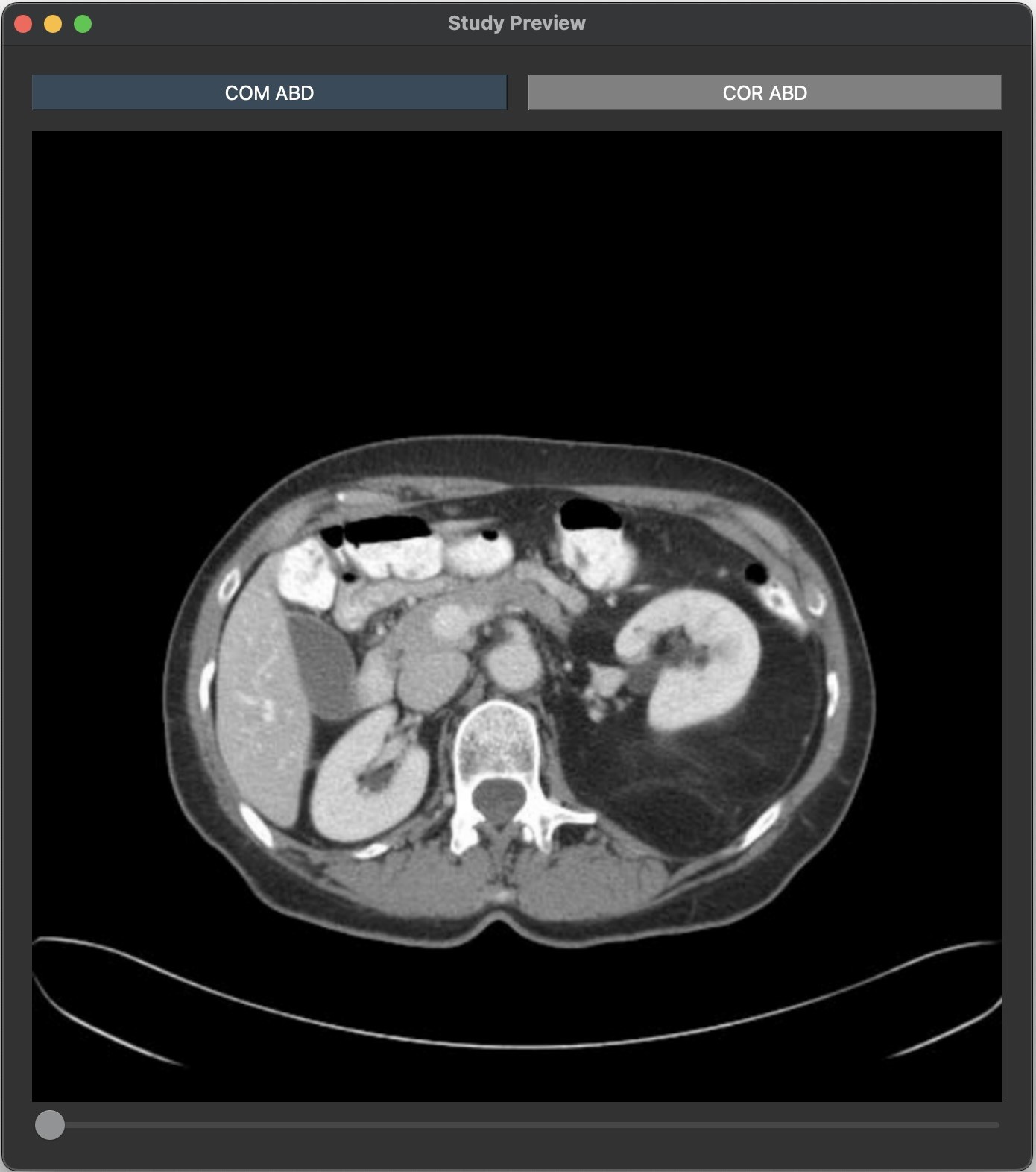}
\caption{Screenshot of the preview dialogue window shown to users for a specific study with two series.}\label{fig3}
\end{figure}

\subsubsection{Settings}
Additionally, the \textit{Settings} view and \textit{Controller} class play a vital role in managing user interactions and backend processes related to PACS configurations and user preferences in the application. The view is initialized by default in the \textit{Connections} page where users can configure and manage network connections to the PACS stations. It is filled with previously saved data from the application settings (stored using the \textit{QSettings} class), which populates a table with the stations’ information, such as the \textit{AE Title}, \textit{IP address}, port, and a descriptive name for the station. This initialization feature is essential, as it provides users immediate access to already configured PACS stations, thus saving time and effort in re-entering information. Additionally, users can perform various actions, including adding new PACS nodes, verifying connections, and removing existing stations.

As for the \textit{Preferences} page, it allows users to customize a variety of settings, such as the ability to get only exact matches for the query contents, adjust timeout settings, and define the path where to save the images.

\subsection{API Testing and Validation}
The API underwent a series of tests during its development, including some major relevant types of API testing \cite{bib15}. Unit testing was conducted to verify the functionality of individual components, with each module tested in isolation to confirm that it met all the requirements for those features. Integration testing was also performed, focusing on how well different components of the API worked together. This process involved combining multiple modules to ensure that data flowed correctly between them and that they interacted as expected. Furthermore, performance testing was also performed to assess the API's responsiveness and stability under varying conditions, which included measuring response times and resource utilization under normal operational loads, ensuring that the API could handle typical user demands efficiently. Additionally, load testing was conducted to simulate scenarios that would require higher processing times, like when retrieving large studies, which was critical to understand if it could affect user experience. Lastly, validation testing ensured that the API met all specified requirements and was fit for its intended purpose. This included confirming that all functionalities worked as intended and that the API delivered accurate results in line with user expectations.

Moreover, the API was tested by a group of individuals including professionals from the healthcare sector, such as radiologists, radiology technicians, and nurses, to obtain feedback on its functionalities and performance, providing valuable insights based on their day-to-day experiences and specific needs in medical imaging workflows. In addition to healthcare practitioners, medical imaging researchers participated in the testing process, whose input helped identify which functionalities could be useful in their experience, while IT professionals also contributed to evaluating the API’s performance. This testing approach was particularly helpful to identify some enhanced features for the application. A questionnaire was developed and distributed among the participants, which aimed to gather both quantitative and qualitative feedback regarding various aspects of the API, such as its usability, performance, and overall functionality, with key areas of focus including ease of navigation, performance, and overall satisfaction with the functionalities offered.

\section{Results and Discussion}
The results of the questionnaire highlight the strong performance of the PACS Browser API in addressing user needs, particularly regarding search functionality and image handling. Users highly valued the flexibility and efficiency of the search features, with specific praise for searching by patient name, custom query fields, and specific series within studies. These functionalities were rated exceptionally well, suggesting they are central to the API's perceived utility. While demographic-specific filters like sex and date of birth were rated slightly lower, the ability to query multiple PACS stations simultaneously received strong support, emphasizing the API’s strength in overcoming fragmentation and ensuring transversality. Image visualization and navigation capabilities also garnered positive feedback. Participants appreciated the clarity of image display, ease of navigation, and the option to convert DICOM images to more accessible formats like JPEG. These features were recognized for their contribution to seamless workflows, facilitating efficient access and management of imaging data across different formats. Performance metrics such as querying and retrieval times were well-rated, reflecting the API's efficiency in delivering timely responses. However, some users noted that initial testing times for connecting to available stations could be improved. Despite this, the overall speed and responsiveness of the system were deemed satisfactory, underscoring its reliability in managing high volumes of medical imaging data. Furthermore, feedback on reliability and task status indication revealed that while users trusted the accuracy of the data retrieved and found the system generally reliable, the task status indication feature was identified as an area needing improvement. Enhancing this functionality could further improve user experience by providing clearer real-time feedback during complex operations.

Overall, the API was seen as a valuable tool for improving workflow efficiency, with users appreciating its intuitive design, ease of use, and effective search and image-handling capabilities. Open-ended feedback further emphasized the API’s strengths, such as its clean interface and comprehensive search options. However, users also identified areas for potential enhancement, including visual design improvements, integrated report creation features, direct access to an image viewer, and the ability to preview images without retrieval. These suggestions provide a clear direction for future development, offering opportunities to refine the API’s functionality and better meet the needs of medical imaging professionals. By addressing these areas, the API could continue to enhance user satisfaction and maintain its effectiveness as a reliable tool in clinical settings.

\section{Conclusions}
The results indicate that the \textit{Transversal PACS Browser API} successfully addresses key challenges faced by healthcare professionals in managing medical images. The application improves usability, efficiency, and interoperability by providing a unified interface for querying and retrieving DICOM images from multiple PACS systems. This approach simplifies workflows, reduces cognitive load, and enhances the user experience. Customizable search functionalities and predefined filters allow users to refine queries more effectively, facilitating faster access to relevant images and reducing delays caused by fragmented systems and data overload. Additionally, the API's ability to convert DICOM images to JPEG and provide image previews further promotes accessibility and interoperability, enabling seamless integration into diverse clinical workflows. Intuitive interface design supports healthcare professionals in navigating large datasets efficiently, while cross-platform compatibility ensures the application works smoothly across different operating systems and devices. Furthermore, the API incorporates security measures to ensure compliance with healthcare regulations such as General Data Protection Regulation (GDPR) and the Portuguese Data Protection Law. 

\section{Future Work}
The proposed API offers significant opportunities for enhancement to better meet user needs in medical imaging. Key improvements include strengthening security with a stronger authentication mechanism, tailoring features for diverse user roles, and increasing flexibility in image retrieval by, for example, enabling users to choose between \textit{C-MOVE} and \textit{C-GET} operations. Addressing the sorting of image slices during DICOM image conversion using reliable attributes like \textit{Image Position Patient} could improve accuracy.Transitioning to a cloud-based deployment could enhance accessibility, scalability, and collaboration among healthcare professionals while accommodating the growing volume of medical imaging data \cite{bib16,bib17}. Additionally, user feedback suggests adding features such as report creation and enhancing status indications for ongoing tasks to streamline workflows.

\begin{credits}
\subsubsection{\discintname}
The authors have no competing interests to declare that are
relevant to the content of this article.
\end{credits}

\bibliographystyle{splncs04}
\bibliography{bib}

\end{document}